\documentclass[superscriptaddress,nofootinbib]{revtex4-1}
\usepackage[utf8]{inputenc}
\usepackage{bm}

\usepackage[dvipsnames]{xcolor}
\definecolor{red}{rgb}{0.9, 0,0}
\definecolor{cerulean}{rgb}{0., 0.42,0.9}
\definecolor{navy}{rgb}{0.05, 0.05,0.8}

\usepackage[colorlinks]{hyperref}
\hypersetup{
  colorlinks = true,
  citecolor = red,
	linkcolor = navy
}
\usepackage{comment}
\usepackage{caption}
\usepackage{subcaption}
\usepackage{bbm}
\usepackage{amsfonts}
\usepackage{mathrsfs}
\usepackage{bbold}
\usepackage{amsmath,amssymb}
\usepackage{subcaption}
\usepackage{siunitx}
\usepackage{slashed}

\usepackage{epsfig}
\usepackage{graphicx}   
\usepackage{url}
\usepackage{float}
\usepackage{soul}
\usepackage{pstricks}
\usepackage{color}
\usepackage{multirow}

\newcommand{\be}{\begin{equation}}
\newcommand{\ee}{\end{equation}}
\newcommand{\bea}{\begin{eqnarray}}
\newcommand{\eea}{\end{eqnarray}}
\newcommand{\beq}{\begin{eqnarray}}
\newcommand{\eeq}{\end{eqnarray}}
\def\bit{\begin{itemize}}
\def\eit{\end{itemize}}
\def\ben{\begin{enumerate}}
\def\een{\end{enumerate}}

\newcommand\DN[1][\relax]{%
\ifx\relax#1\relax\else{}^{#1}\fi \!X}

\DeclareMathAlphabet\mathbfcal{OMS}{cmsy}{b}{n}

\definecolor{cerulean}{rgb}{0., 0.52,0.65}

\graphicspath{{plots/}}

\begin{document}
\title{Calorimetric Detection of Dark Matter}
\author{Julien Billard}
\affiliation{Univ. de Lyon, Université Lyon 1, CNRS/IN2P3, IPN-Lyon, F-69622 Villeurbanne, France}
\author{Matt Pyle}
\affiliation{Department of Physics, University of California, Berkeley, CA 94720}
\author{Surjeet Rajendran}
\affiliation{Department of Physics and Astronomy, The Johns Hopkins University, Baltimore, MD 21218, USA}
\author{Harikrishnan Ramani}
\affiliation{Stanford Institute for Theoretical Physics,
Stanford University, Stanford, CA 94305, USA}
\begin{abstract}
Dark matter direct detection experiments are designed to look for the scattering of dark matter particles that are assumed to move with virial velocities  $\sim 10^{-3}$. At these velocities, the energy deposition in the detector is large enough to cause ionization/scintillation, forming the primary class of signatures looked for in such experiments. These experiments are blind to a large class of dark matter models where the dark matter has relatively large scattering cross-sections with the standard model, resulting in the dark matter undergoing multiple scattering  with the atmosphere and the rock overburden, and thus slowing down considerably before arriving at underground detectors. We propose to search for these kinds of dark matter by looking for the anomalous heating of a well shielded and sensitive calorimeter. In this detector concept, the dark matter is thermalized with the rock overburden but is able to pierce through the thermal shields of the detector causing anomalous heating. Using the technologies under development for EDELWEISS and SuperCDMS, we estimate the sensitivity of such a calorimetric detector. In addition to models with large dark matter - standard model interactions, these detectors also have the ability to probe dark photon dark matter. 
\end{abstract}
\maketitle

\section{Introduction}
\label{sec:intro}

Dark matter (DM) direct detection efforts are concentrated on dark matter particles whose scattering cross-sections with the Standard Model (SM) are small enough that the dark matter travels unhindered to the detector and interacts with the detector with a velocity equal to the galactic virial velocity $\sim 10^{-3}$. This assumption informs the experimental design of direct detection experiments - these are located in underground mines (to minimize cosmic ray backgrounds) and search for dark matter energy depositions in the keV range. This assumption is true in a large class of models. In light of null results from dark matter direct detection experiments, it is important to question this assumption. There are generic, theoretically well motivated models where this assumption is not valid. These models include strongly (i.e. QCD) interacting dark matter and composite dark matter. In both these cases, the dark matter - standard model scattering cross-section can be larger than $\sim 10^{-29} \text{ cm}^2$ causing the dark matter to scatter multiple times before it can reach the $\sim$ km depths of underground detectors. This scattering causes the dark matter to slow well below its virial velocity, leading to suppressed energy deposition in direct detection experiments. 

Strongly interacting dark matter can be constrained from energy deposition in rocket~\cite{mccammon2002high}, balloon~\cite{rich1987search} and other surface experiments~\cite{CRESST:2017ues}. While these preclude the case that all of the dark matter is strongly interacting, due to their reduced exposure, there are few constraints on a sub-component of strongly interacting dark matter. It is important to constrain this scenario since generic models of strongly interacting dark matter do produce scenarios where the majority of the dark matter is in a deep QCD bound state with reduced cross-section but with a small abundance of hadronized states that have QCD scale scattering cross-sections. This includes models of ``colored dark matter'' where the dark matter arises from stable color octet dirac fermions \cite{DeLuca:2018mzn}. Another viable scenario is mediated by a dark photon in the 10s of MeV mass range~\cite{Berlin:2021zbv,McKeen:2022poo}

Richer phenomenology is possible in composite DM models. In these models where the dark matter is an agglomerate with a very large number of particles \cite{Rajendran:2020tmw, Hardy:2014mqa, Wise:2014jva, Grabowska:2018lnd}, low momentum scattering is coherently enhanced while high momentum scattering is suppressed. In these models,  hard collisions are highly suppressed and are not detectable in low exposure experiments, permitting all of the dark matter to be in these composite states \cite{Rajendran:2020tmw}. 

How can these models be probed? A generic feature of their phenomenology is the accumulation of such dark matter in the earth, leading to a trapped but slow dark matter density that can be many orders of magnitude larger than the galactic density. This trapped population can arise entirely from scattering with the SM. But it may also emerge when only a small fraction of the dark matter is stopped by the earth - in this case, depending upon the details of the dark matter model, self interactions in the dark sector can cause a trapped initial seed populaton of dark matter to slow down infalling dark matter, causing the trapped population to grow over time~\cite{Rajendran:2020tmw}. This trapped population thermalizes with the earth and is distributed over the entire volume of the earth for $\sim$ GeV mass dark matter \cite{Neufeld:2018slx} with the trapped population sinking towards the earth's core for higher masses. But even in the latter case, the slow sinking of the dark matter leads to a ``traffic jam'' \cite{Pospelov:2019vuf} as the dark matter collides its way down to the core leading to an enhanced local density. The key challenge in probing this kind of dark matter is its low kinetic energy.

One way to probe these kinds of dark matter is to use calorimetry. The key idea is that the trapped dark matter is in thermal equilibrium with the environment. Since the dark matter scattering cross-sections are low, this temperature is set by the environment that is at 300K, irrespective of shielding.  Consider the following setup: place a material that is well isolated from the environment and attach a calorimeter to it.  The dark matter pierces the thermal shield and heats the material, leading to anomalous heating.  This effect was used to constrain dark matter in \cite{Neufeld:2018slx} by using the observed lifetime of cold Helium dewars. Their limit $\sim 10 \mu\text{W/mole}$ already limited interesting parameter space even though the setup was not optimized to look for this signal. Moreover, it is likely that existing data from highly sensitive calorimetric detectors such as CRESST, EDELWEISS, and SuperCDMS, can place better limits on dark matter parameter space than those placed by \cite{Neufeld:2018slx}. But, such a limit would require  careful reanalyses of the data. 

While such  reanalyses could be useful, it is interesting to estimate the scientific reach of an optimized setup that is dedicated to search for such anomalous heating. This optimized setup is likely to be a far more potent probe of this class of dark matter models rather than constraints placed from reanalyses of existing data.  In this paper, we focus on the highly sensitive calorimetric techniques developed by the CRESST, EDELWEISS, and SuperCDMS collaborations and investigate if they can be deployed to look for such anomalous heating. It is also interesting to ask if these techniques, when pushed to their technological extreme, can search for other dark matter candidates such as light hidden photon dark matter which have a large abundance and can be absorbed in materials. In this paper we develop this concept. In section \ref{sec:concept} we present a conceptual overview of the detector and highlight the backgrounds that need to be overcome. The technical implementations of the concept are discussed in section \ref{sec:technical}. We describe the sensitivity of this setup to various dark matter models in section \ref{sec:models}. We then conclude in section \ref{sec:conclusions}. 

\section{Concept}
\label{sec:concept}

In the following section, we introduce the basic concept of a dedicated cryogenic detector to search for anomalous heating. We also make some simple order of magnitude estimates of the sensitivity of such a detector. A more detailed discussion of these estimates is made in section \ref{sec:technical}. 

An efficient detection of heating power from external sources requires the following three basic principles. First, the cryogenic detector needs to be maximally decoupled from the environment while retaining optimized sensitivity to constant power excitation. Second, the detector needs to avoid any sources of external and internal parasitic power, for example from elastic deformations, vibrations, electromagnetic interference and particle interactions from radioactive backgrounds. Finally, the detector needs to operate in stable conditions with ways to turn the dark matter (DM) signal ON and OFF with an accurate control of the systematics. This final principle is particularly important since unlike conventional WIMP calorimeters, the detection signals that are being searched for in this experiment do not cause ionization or scintillation and thus lack conventional handles that are used to discriminate the signal from background.  

\begin{figure}[t!]
\includegraphics[width=0.99\textwidth,angle=0]{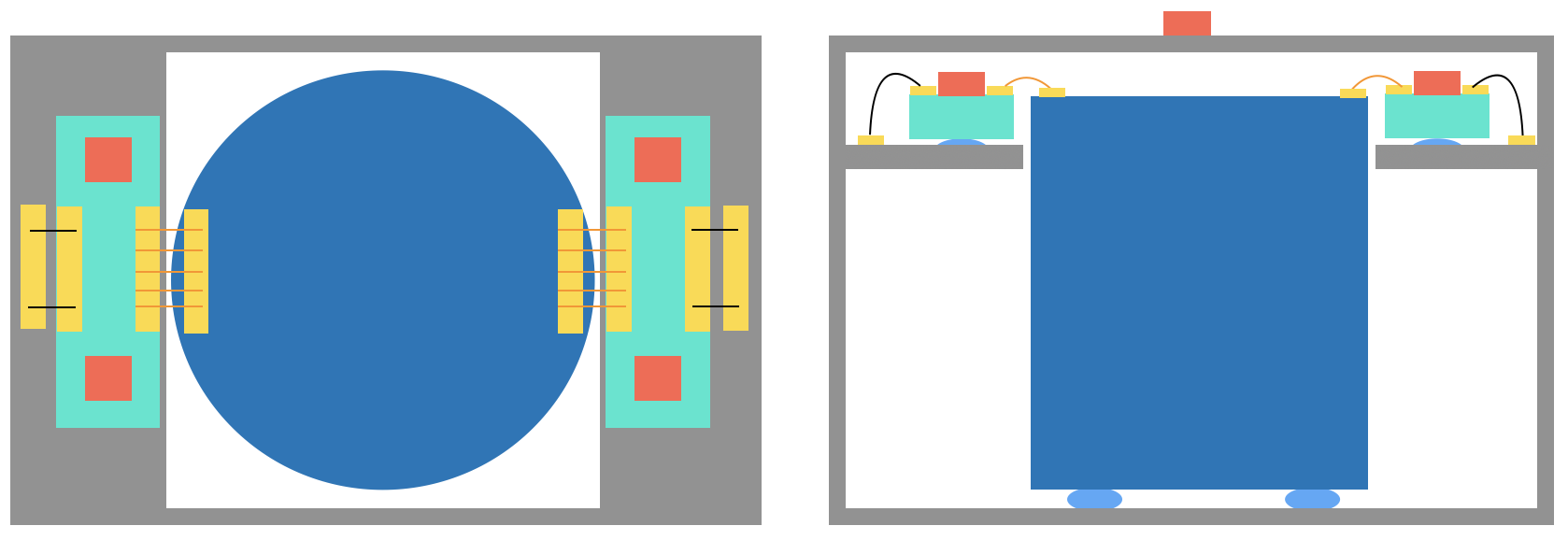}
\caption{ Top (left) and side (right) views of our proposed cryogenic detector design concept with: a 20 kg target crystal (blue), two sapphire slabs (light green) each equipped with two NTD heat sensors (red) immersed into a large 10 mK thermal screen (grey). The thermal connections are done thanks to gold pads (yellow) and either gold (orange) or aluminum (black) wirebonds. The sapphire slabs and the large target crystal are efficiently decoupled from the 10 mK thermal stage by mechanically resting on insulating sapphire spheres or slabs (cyan).} 
\label{fig:BasicConcept}
\vspace{-0.5cm}
\end{figure}

The general idea of our detector concept shown in Fig.~\ref{fig:BasicConcept} is the following:
\begin{itemize}
    \item A large target crystal (blue) is resting on low-thermal conductance clamps (cyan) maximizing its thermal decoupling from the cryostat. The corresponding thermal conductance is called $G_{ab}$.
    \item The heating power is read through the use of two sensor chips (light green) each equipped with two germanium Neutron Transmutation Doped sensors~\cite{Mathimalar:2014sfa} (NTD - shown in red) thermally coupled to the target crystal with a thermal conductivity of $G_{as}$. The latter is maximized by the use of large gold pads (yellow) and gold wirebonds (orange). Note that the use of two sensing chips (with a total of four NTD) allows to reject correlated noise.
    \item The thermal connection from the sensing chips to the cryostat ($G_{sb}$) is tunable, thanks to the use of either gold or aluminum wirebonds (black) and varying gold pad sizes, and designed to ensure a maximal sensitivity to any DM heating of the crystal.
    \item An additional NTD is mounted directly on the copper housing in order to follow any temperature drift of the system with high accuracy.
\end{itemize}
An optimized system must follow the basic concept that $G_{ab} \ll G_{sb} \ll G_{as}$, while considering additional constraints from internal and external parasitic powers to be mitigated and that may require additional tuning of the various thermal conductivities. 

As stated above, the detector design must allow to turn the DM signal ON and OFF in order to properly estimate all the thermal conductivities of the system and to assess the sensitivity to any DM induced external heating of the crystal. To that order, this anomalous heating search must be done in a sequence of several measurements. Firstly, we start with no target crystal in order to properly characterize the thermal sensors. As we have two thermal sensors per chip, we can derive with a high level of precision their thermal response as well as the various conductivities at play, including the one to the bath ($G_{sb}$). In a second step, we connect the two sensing chips to the target crystal in order to assess the remaining thermal conductivities of the crystal to the bath ($G_{ab}$) and of the sensing chip to the crystal ($G_{sa}$). By heat loading the system via the NTD of the sensing chips (as in the previous step) and then via illuminating the crystal target with an infrared LED, with a calibrated energy output, one can precisely derive both ($G_{sa}$) and ($G_{ab}$). Lastly, as the DM interaction strength varies with the target material, this setup is designed to be flexible enough to switch to any target material of interest to maximize our DM sensitivity\footnote{Note that to perform each of these steps with maximal reproducibility, and control of systematics, it is of utmost importance that the detector design allows for all these modifications to be made with minimal changes ({\it e.g.} only bonding or breaking wire bonds linking the target crystal to the sensing chips)}. Thanks to this sequence of measurements, with and without target crystal and varying the target materials, we expect to be able to properly subtract any sources of parasitic power and push our DM sensitivity as close as possible to our ultimate sensitivity to external DC power heating.

Following the detailed calculations discussed in the next section (Sec.~\ref{sec:technical}) we find that considering an idealized 20 kg Ge crystal, 10 mK temperature, and no radioactive background, a DC power heating sensitivity of $1.9\times10^{-21}$~W can be achieved in only one week of DM search data with NTD sensors. More realistically, considering a 1~kg Ge target crystal and a low-background environment as measured by the EDELWEISS-III experiment in Modane~\cite{Armengaud:2017rzu}, an upper limit of $1\times10^{-20}$~W can be achieved in only two weeks of science run. Considering instead low-impedance Transition Edge Sensors (TES), with critical temperatures of 20~mK, should lead to an improved sensitivity up to about one order of magnitude especially in the presence of background requiring fast time response, suggesting that sensitivities to any DM induced anomalous heating of the order of $10^{-22}$~W , and even lower with lower background and larger exposition time, could be achievable with our detector concept. As further detailed in Sec.~\ref{sec:models} this opens up the possibility to probe a plethora of DM scenarios that would escape more traditional detection methods.

\section{Technical Implementation}
\label{sec:technical}

We hereafter present the thermal modelisation of the proposed cryogenic detector design presented in Sec.~\ref{sec:concept}. For the sake of concreteness, the following calculations are presented in the context of the use of NTDs as thermal sensors, but similar equations and methods can be translated straightforwardly to the case of low impedance TES.

The electro-thermal modelisation of the detector design presented above, specifically tuned to search for external heating power sources, is characterized by a set of non-linear differential equations describing the heat flow between the different heat capacities and their couplings to the electronic readout. We consider the following convention describing the heat flow between two systems at temperature $T_x$ and $T_y$ as $P_{xy} = g_{xy}(T_x^n - T_y^n)$ with a thermal conductivity $G_{xy} = dP_{xy}/dT = ng_{xy}T^{n-1}$, where $g_{xy}$ is the thermal coupling constant.

As illustrated in Fig.~\ref{fig:BasicConcept}, the system consists of a total of four NTDs, two sapphire slabs, and a massive target crystal with gold and aluminum wirebonds, and sapphire balls (or slabs) ensuring the thermal contacts between these elements. Except for NTDs, which have both an electron and phonon system thermally connected to each other via their electron-phonon coupling ($g_{ep}$), the sapphire slabs and the target crystal are only characterized by their phonon heat capacity. The system is therefore composed of 11 thermal and 4 electrical non-linear differential equations which, following the methodology from~\cite{Billard:2016apk}. They can be written as :

\begin{align}
\label{eq:ODE}
C_{c_{i,j}} \frac{d V_{i,j}}{d t} &= \frac{V_{b_{i,j}} - V_{i,j}}{R_L} - \frac{V}{R(T_{e_{i,j}})} 
\end{align}
\begin{align}
C_{e_{i,j}} \frac{d T_{e_{i,j}}}{d t} &= \frac{V_{i,j}^2}{R(T_{e_{i,j}})} - V_{NTD} g_{ep} \left( T_{e_{i,j}}^{n} - T_{p_{i,j}}^{n} \right) \nonumber
\end{align}
\begin{align}
C_{p_{i,j}} \frac{d T_{p_{i,j}}}{d t} &= V_{NTD} g_{ep} \left( T_{e_{i,j}}^{n} - T_{p_{i,j}}^{n} \right) - g_{glue} S_{NTD} \left( T_{p_{i,j}}^{n_g} - T_{s_{i}}^{n_g} \right) \nonumber
\end{align}
\begin{align} 
C_{s_i} \frac{d T_{s_{i}}}{d t} = g_{glue} S_{NTD} \left[ \left( T_{p_{i,1}}^{n_g} - T_{s_{i}}^{n_g} \right) + \left( T_{p_{i,2}}^{n_g} - T_{s_{i}}^{n_g} \right) \right] + g_k S_{AuGe_i}\left( T_a^{n_k} - T_{s_i}^{n_k} \right) - g_k S_{AuB_i}\left(T_b^{n_k} - T_{s_i}^{n_k}\right) \nonumber
\end{align}
\begin{align}
C_a \frac{d T_{a}}{d t} &= - \left[ g_k S_{AuGe_a}\left(T_a^{n_k} - T_{s_a}^{n_k}\right) + g_k S_{AuGe_b}\left(T_a^{n_k} - T_{s_b}^{n_k}\right) + g_b N_{Ge_B}\left(T_b^{n_b} - T_{a}^{n_b}\right) \right] \nonumber
\end{align}

where $V_{i,j}$ is the NTD output voltage with its cabling capacitance $C_{c_{i,j}}$, $T_{e_{i,j}}$ and $T_{p_{i,j}}$ are the temperatures of the electron and phonon systems of the NTDs and their corresponding heat capacities $C_{e_{i,j}}$ and $C_{p_{i,j}}$, $T_{s_{i}}$ is the temperature of the sapphire slabs and $C_{s_i}$ their heat capacity, $T_a$ is the temperature of the target absorber crystal of heat capacity $C_a$, and $T_b$ is the temperature of the detector holder (also called bath temperature). The indices $i$ and $j$ respectively refer to the sapphire slabs $\{a, b\}$ and the NTDs $\{1,2\}$ of each slabs. $V_{b_{i,j}}$ and $R_{i,j}$ are respectively the NTD bias and resistance. The latter depends on the electron system temperature $T_e$ following the Shklovskii-Efr\"os law:  $R(T_e) = R_0\exp(\sqrt{T_0/T_e})$ where $R_0$ depends on the intrinsic properties of Ge and of geometrical factors, and $T_0$ is related to the germanium doping level~\cite{Mathimalar:2014sfa}. The various $n$ exponents correspond to the power law of the different heat conductances. $V_{NTD}$ and $g_{ep}$ are the volume and electron-phonon coupling constants of the NTD. $S_{NTD}$ and $g_{glue}$ are the NTD surface, {\it i.e.} the gluing surface, and the glue coupling constant\footnote{For a review of NTD electron-phonon and glue coupling constant measurements, see~\cite{novati:tel-01963790} and reference therein.}. $S_{AuGe}$, $S_{AuB}$ and $g_k$ are the surfaces of the gold pads on the sapphire slabs connecting to the absorber and to the detector holder respectively, and their corresponding Kapitza conductances~\cite{marnieros:tel-01088881}. Lastly, $N_{Ge_B}$ and $g_b$ are respectively the number of sapphire balls holding the target crystal and their thermal conductivity~\cite{Pinckney:2021wtd}.\\

From Eq.~\ref{eq:ODE}, one can fully describe the detector response in all three regimes of interest: 1) the steady state when everything is in equilibrium, 2) the time domain where we can simulate pulses arising from particle interactions in the target crystal or any other component of the system, and 3) the frequency domain which is used to study the noise power spectral densities and related energy resolutions and performance. 

\begin{figure}[t!]
\includegraphics[width=0.9\textwidth,angle=0]{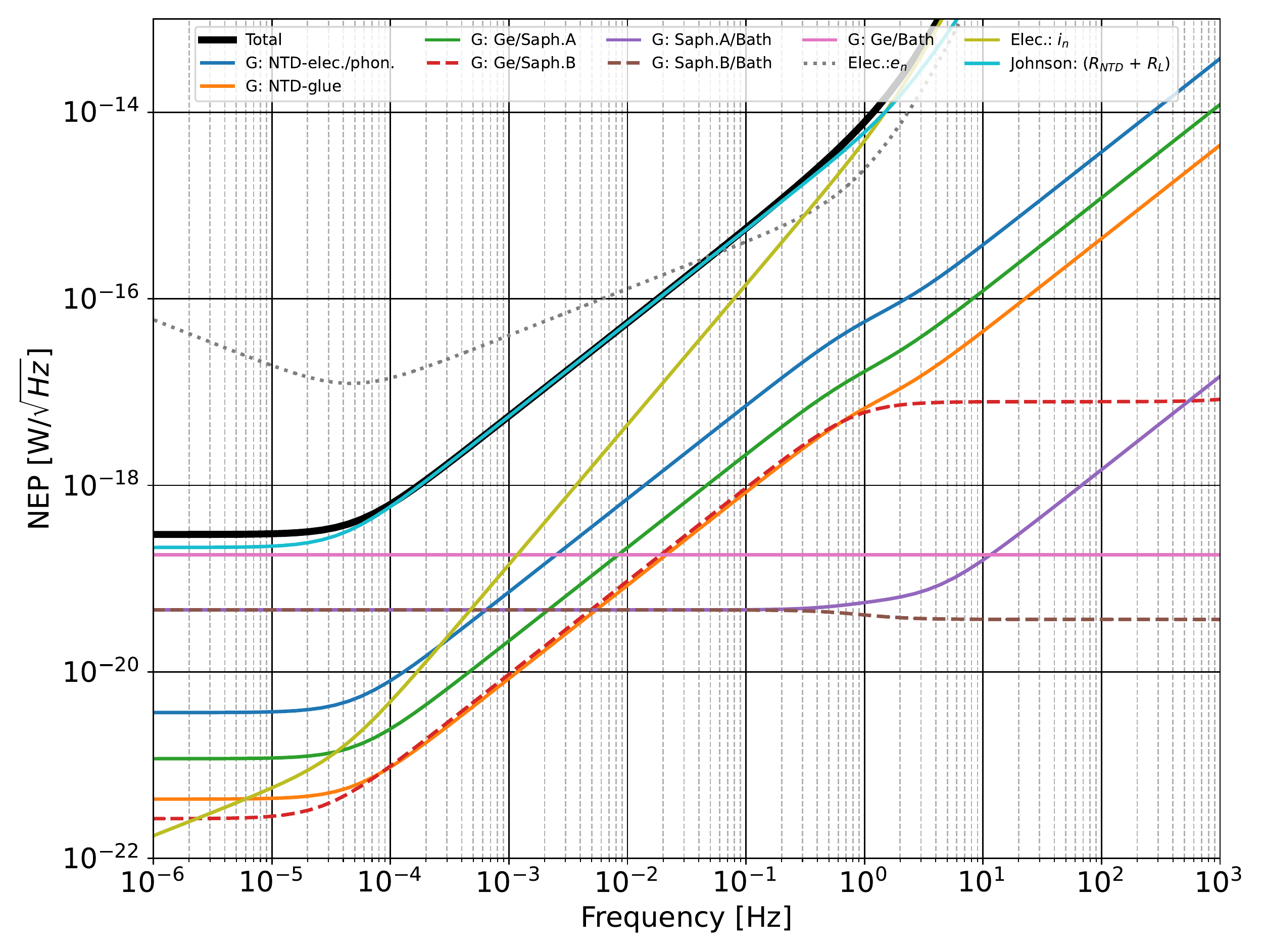}
\caption{Noise Equivalent Power from the dominating and most relevant sources as a function of frequency for one of the four NTDs, namely ($i=a$,$j=1$). The computations have been done for an optimized NTD size of 20x20x1~mm$^3$, a bias voltage of 11~mV, a detector holder temperature of 10~mK, a 20~kg Ge target crystal and the parameter values given in Tab~\ref{tab:parameter_IdealDetector}}
\label{fig:NEP_IdealDetector}
\vspace{-0.5cm}
\end{figure}

The total noise power spectral density referenced to the output voltage of the preamplifier $S_{V,tot}$ is computed in summing quadratically all sources of noise, assuming that they are uncorrelated. We expect three distinct sources of noise: 1) thermal fluctuation noise (TFN), 2) current noise from the preamplifier and the resistors, and 3) voltage noise from the preamplifier.\\
Thermal fluctuation noise arise at each thermal link ($G_{ij}$) so that the total TFN contribution referenced at the output voltage can be expressed as~\cite{Galeazzi:2003pe,Irwin:1995fx}:
\begin{equation}
S_{V,TFN} = 2k_B \left[ \sum(T_i^2 + T_j^2) G_{ij} \left\vert Z_{vi}^{-1} - Z_{vj}^{-1} \right\vert^2 \right] \qquad [V^2/Hz]
\end{equation}
where the indices $i$ and $j$ are referring to two systems thermally connected to each other. $k_B$ is the Boltzmann constant and $Z_{vi}$ is the element from the electro-thermal impedance matrix, derived from the Fourier transform of Eq.~\ref{eq:ODE}~\cite{Galeazzi:2003pe,Irwin:1995fx}, connecting the $i$-th thermal system to the considered output NTD voltage $v$. \\
The total current noise, coming from the preamplifier ($i_n$) and the Johnson noise contributions from the NTD thermistance $R(\bar{T}_e)$ and the load resistor $R_L$ heat sunk at $T_{R_L}$ is:
\begin{equation}
S_{V,i} = \left[ i_{n}^2 + 4k_B\left( \frac{\bar{T}_e}{R(\bar{T}_e)} + \frac{T_{R_L}}{R_L} \right) \right] \left\vert Z_{vv}^{-1}\right\vert^2 \quad [V^2/Hz]
\end{equation}
Finally, the last noise contribution comes from the voltage noise of the preamplifier $e_n$ which is directly referenced to the voltage output as it is independent of the impedance such that:
\begin{equation}
S_{V,e} = e_n^2 \quad [V^2/Hz]
\end{equation}
where the voltage ($e_n$) and current ($i_n$) noise from the preamplifier are taken from~\cite{RICOCHET:2021fix} considering the case of a 200~pF High Electron Mobility Transistor (HEMT). 
The Noise Equivalent Power (NEP), which corresponds to the amount of thermal noise in the absorber that would produce an identical amount of voltage noise in the NTD, is defined as~\cite{Irwin:1995fx,Galeazzi:2003pe,Pyle:2015pya},
\begin{equation}
\label{eq:nep}
{\rm NEP}^2(\omega) =  \frac{S_{V,tot}}{|\hat{s}_V|^2} \qquad [W^2/Hz]
\end{equation}
where $\hat{s}_V$ is the detector signal sensitivity referenced to one of the four NTD output voltage following:
\begin{equation}
    s_V(\omega) = Z_{va}^{-1} \qquad [V/W]
\end{equation}
assuming that all of the phonons produced in the target crystal are instantaneously thermalized. Eventually, the DC power energy resolution is given by:
\begin{equation}
\label{eq:DCpower}
\displaystyle{ \sigma_{\rm DC}^2 = \int_{0}^{\omega_0} \frac{\mathrm{d}\omega}{2\pi}|{\rm NEP}(\omega)|^2}  \qquad [W^2]
\end{equation}
with $\omega_0 = 2\pi/t_0$ where $t_0$ the experiment's exposition time. In the case that the NEP is constant down to 0~Hz, and as long as the experiment is not limited by any other sources of parasitic power, its sensitivity to a DC external power heating scales as $\sigma_{\rm DC} = {\rm NEP(\omega=0)} / \sqrt{t_0}$.

\begin{figure}[htpb]
\includegraphics[width=0.8\textwidth,angle=0]{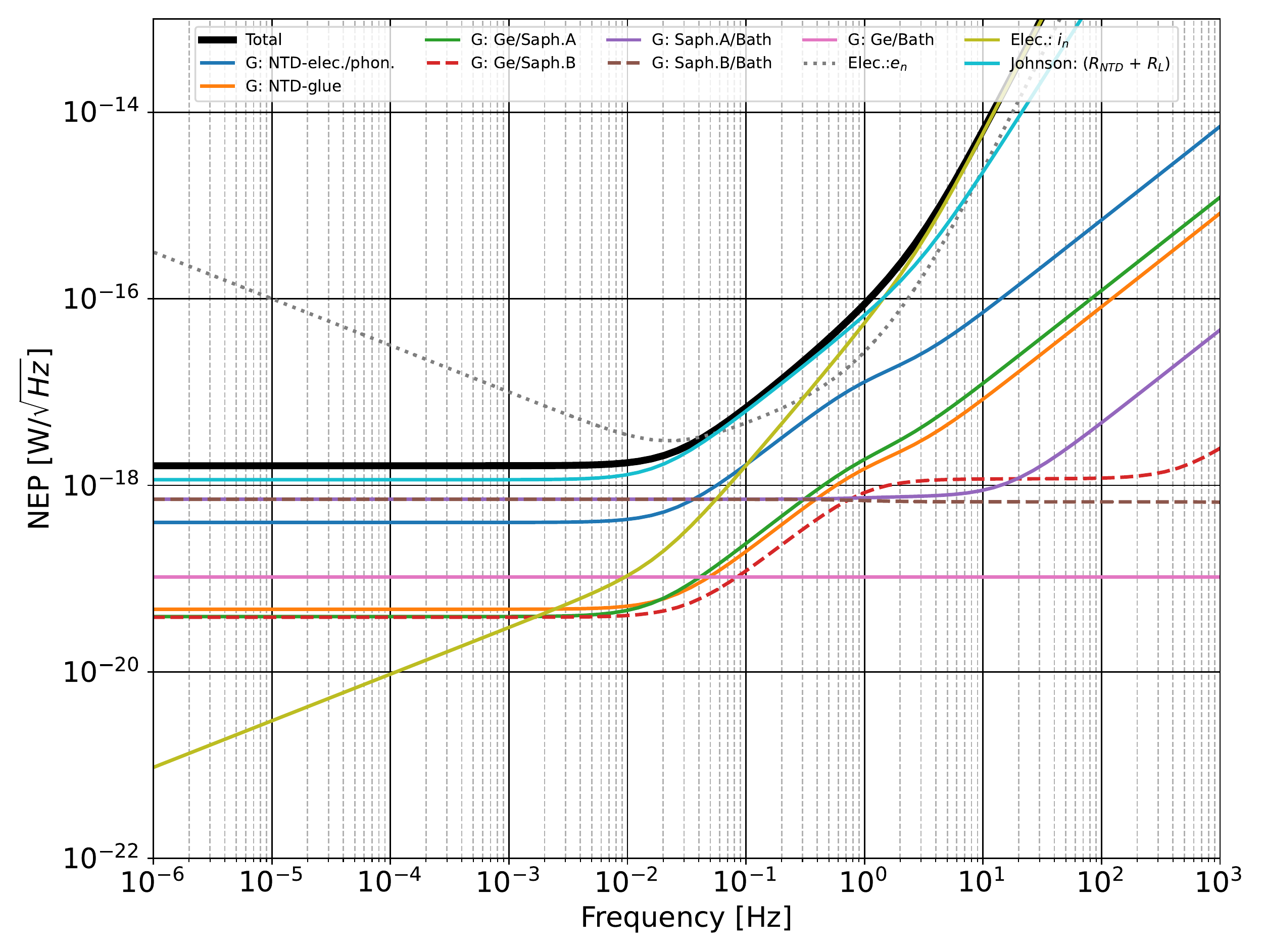}
\includegraphics[width=0.8\textwidth,angle=0]{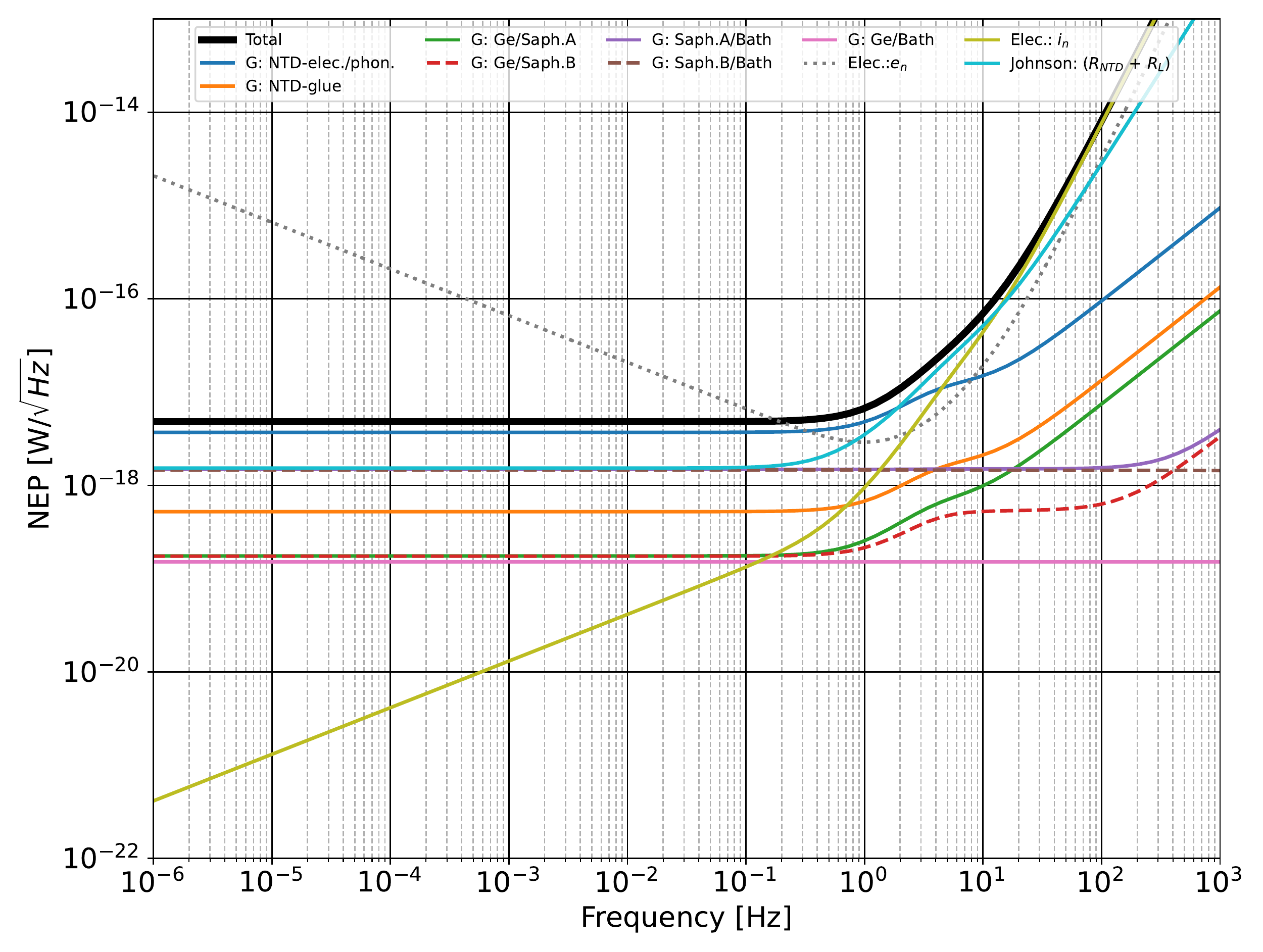}
\caption{Noise Equivalent Power from the dominating and most relevant sources as a function of frequency for one of the four NTDs, namely ($i=a$,$j=1$) for a 1 kg Ge crystal operated underground (left panel) and a 40~g crystal operated above ground (right panel). The model parameter values are given in Tab~\ref{tab:parameter_1kgDetector} and Tab~\ref{tab:parameter_40gDetector}, respectively.}
\label{fig:NEP_RealDetectors}
\vspace{-0.5cm}
\end{figure}

We hereafter use and optimize our model parameters in the context of three different scenarios in order to estimate their anomalous heating power sensitivities\footnote{for the sake of concreteness we consider a Ge target material, but similar calculations can be done with any other type of materials}: 1) an idealized 20 kg Ge detector without radioactive backgrounds, 2) a 1 kg Ge crystal exposed to a radiogenic background as observed by the EDELWEISS-III experiment in Modane~\cite{Armengaud:2017rzu}, and 3) a 40~g Ge crystal operated above ground with a radioactive background environment similar to the one observed by the EDELWEISS-Surf experiment~\cite{EDELWEISS:2019vjv}.
\begin{itemize}
    \item {\bf Idealized 20 kg Ge target crystal:} Assuming no backgrounds we only constrained our thermal model optimization to achieve a decaying time constant of about an hour ($\tau_d = 3417$ ~s). Increasing this time constant will improve the overall sensitivity but at the cost of very long time periods dedicated to the measurements of the various thermal conductivities which could jeopardize the feasibility of this detector operation. Table~\ref{tab:parameter_IdealDetector} gives the numerical values of the corresponding model parameters. Using very large $20\times20\times1 = 200$~mm$^3$ NTDs and a $G_{AuB} = 3.6\times 10^{-13}$~W/K thermal conductivity between the sapphire slabs and the bath, and further assuming the use of an AC modulating readout electronics, as done in EDELWEISS-III to reject the voltage noise from the preamplifier ($e_n$), a DC power sensitivity at the 5-$\sigma$ level of $1.86\times 10^{-21}$~W is achieved after one week of science data taking. Figure~\ref{fig:NEP_IdealDetector} presents the total NEP at the NTD \{a,1\} as well as the dominating and/or the the most relevant individual sources of noise. One can then appreciate that the detector's sensitivity is limited by the Johnson noise from the 947~M$\Omega$ NTD impedance and the thermal coupling between the target crystal and the bath, suggesting that our thermal design is well optimized. Interestingly, we find similar sensitivities when using a low-impedance TES instead of an NTD in this ideal case where there is no strong constraint on the detector's time response to particle interactions.
    \item {\bf Underground 1 kg Ge target crystal:} Considering a more realistic scenario including radioactive backgrounds, we now consider the one observed by the EDELWEISS-III experiment operated at the Laboratoire Souterrain de Modane (LSM) in their low-background environment. The EDELWEISS-III experiment has reported an event rate in their 860~g Ge detectors of about 4.3~mHz  with a mean energy deposition of 500~keV~\cite{Armengaud:2017rzu}. Assuming a 1~kg Ge crystal hereafter, this leads to an event rate of about $r=$~5~mHz corresponding to a resulting radioactive parasitic power of $4\times 10^{-16}$~W. This background can however be efficiently mitigated with a time response of the detector to particle interaction fast enough to actively reject or subtract these energy depositions. In the thermal model parameter optimization considered hereafter, we tuned the thermal leak of the system ($G_{AuB}$) to reach a decaying time constant satisfying $\tau_d<0.05/r = 10$~s, ensuring an averaged DC livetime of about 50\% in rejecting pulses over a $10\times \tau_d$ time window. The resulting optimized parameters are given in Tab.~\ref{tab:parameter_1kgDetector} and the resulting NEP is shown in Fig.~\ref{fig:NEP_RealDetectors} (left panel). As one can see, considering a $5\times 5\times 1 = 25$~mm$^3$ NTD, the detector performance is now limited by both the NTD intrinsic Johnson noise and the stronger sapphire slabs-to-bath thermal conductivity to ensure a sufficiently short decay time constant of about 7~s. Eventually, we find that such a detector design can achieve a DC power sensitivity at the 5-$\sigma$ level of $1.04\times 10^{-20}$~W after two weeks of science data taking considering a 50\% livetime efficiency.
    \item {\bf Above ground 40~g Ge target crystal:} Assuming a significantly larger radioactive background as observed by the above ground EDELWEISS-Surf experiment about $10^4$ larger than the one from the EDELWEISS-III experiment~\cite{EDELWEISS:2019vjv}, we consider a 40~g Ge crystal. With an event rate $r = 0.2$~Hz we tuned the thermal model parameters to achieve a decay time constant of the detector $\tau_d = 142$~ms, well below the 250~ms requirement. The resulting optimized parameters, where we now consider $2\times 2\times 1 = 4$~mm$^3$ NTDs, are given in Tab.~\ref{tab:parameter_40gDetector} and the resulting NEP is shown in Fig.~\ref{fig:NEP_RealDetectors} (right panel). The DC power sensitivity of this detector operated above ground is found to be of $3.08\times 10^{-20}$~W after two weeks of science data taking. Interestingly, as the detector needs to have a faster time response, we see that its performance become limited by the weak NTD electron-phonon coupling which was found to be the main limitation of using NTD with respect to low impedance TES.
\end{itemize}

In conclusion of this short study dedicated to the estimation of our detector design to dark matter induced heating power, we found that depending on the radioactive backgrounds, sensitivities down to $10^{-20}$~W (1 kg crystal) are realistically achievable. Interestingly, we find that going from an idealized background-free 20~kg crystal to a 40~g one operated above ground degrades the sensitivity by about four orders of magnitude, still suggesting that interesting parameter space can be explored on a short time scale from above ground setups, see Sec.~\ref{sec:models}. It is important to note that in this study we have only addressed the limitations from radioactive backgrounds and in principle, all sources of constant or quasi-constant parasitic power, such as elastic deformations, vibrations, infrared radiations and others, also need to be maximally mitigated.

\section{The Theory Space} 
\label{sec:models}

In this section, we present an example of  a well motivated large cross-section DM-SM model that can be probed using the above calorimetric detection concept. Our proof of concept is a model where the DM interacts with SM via a 4-fermi contact interaction. While this detection technique is applicable to a wider class of models, including models of composite dark matter ``blobs'', we defer their discussion to future work. This is because, the primary goal of this paper is to demonstrate the key ideas of a calorimetric detector and the scope of the theory section is to demonstrate the existence of at least one class of well motivated theories that can be probed by such a detector. 

As is common in the DM literature, we parameterize the DM-SM interaction through the per-nucleon interaction. Accordingly, the 4 fermi contact operator of interest is parameterized as: 

\begin{align}
\mathcal{L}\supset \frac{1}{\Lambda^2}\bar{n}{n}\bar{\chi}{\chi},
\end{align}
where $n$ is the nucleon and $\chi$ is the dark matter. This yields the per nucleon cross-section  $\sigma_n=\frac{m_n^2}{\pi \Lambda^4}$ when the dark matter mass is bigger than the nucleon mass. As we discuss below, we will be interested in models of dark matter with a mass greater than GeV and thus this limit is appropriate. This 4-fermi operator encompasses a broad class of models. This operator is appropriate whenever the mass of the mediator responsible for DM-SM scattering, $m_V$ is larger than the typical momentum transfers involved in the collisions {\it i.e.} $m_V \ge q_{\rm vir}= 2\mu v_{\rm vir} \approx 50 \textrm{MeV} \gg q_{\rm th} \approx 2\mu v_{\rm th}$. Here $q$ and $v$ stand for momentum and velocity and the subscript ``vir" and ``th" stand for virial and thermal. This could arise in models where the DM contains particles charged under QCD, such as sexaquark dark matter~\cite{Jaffe:1976yi,Farrar:2003gh,Farrar:2005zd,Farrar:2017eqq}, or hybrid hadrons~\cite{DeLuca:2018mzn,Beylin:2020bsz}. In this case, the mediator $V$ can be identified as the SM mesons. Another possibility is that the mediator is a dark photon $A'$ as explored in~\cite{Berlin:2021zbv,McKeen:2022poo}. 

The transfer cross-section $\sigma_T$ relevant to computing the number densities, is given by,
\begin{equation}
    \sigma_T=\textrm{Min}\left[4\pi R_A^2,\frac{A^2\mu^2}{m_n^2}\sigma_n \right]
\end{equation}
Here, $R_A$ is the radius of the SM rock/atmospheric atom and A is its atomic mass, $\mu$, the reduced mass of the atom-DM. The $A^2$ enhancement is replaced by $Z^2$ for dark photon mediated models. Note that the cross-section is cut-off so as to never increases much above the geometric size of the nuclei. 

With this cross-section, we now summarize the estimates of its local abundance and compute the heating rate in the experiment and make projections using our benchmark heating rates. In the parameter space of interest to this experiment, the large DM-SM cross-section causes repeated scattering of the DM in the atmosphere and rock overburden causing the DM to slow down and thermalize before reaching the detector.  If the resultant thermal velocity $v_{\rm th} \approx \sqrt{T_{\rm room} m_{\rm DM}}$ is smaller than the Earth escape velocity $v_{\rm esc}$, then the thermalized DM accumulates over the age of the Earth leading to large local build-up. The mass where $v_{\rm esc}> v_{\rm th}$, roughly corresponds to $m_{\rm DM} \ge \textrm{GeV}$, and thus as mentioned earlier,  we will only consider masses above a GeV.  Whereas masses less than a GeV evaporate, masses greater than a GeV sink to the Earth's core. The resultant static population of DM accessible near the Earth's surface can be computed via solving the Jean's equation which captures the effect of the Earth's gravitational potential as well as the effects of temperature and density variations inside the Earth. This density $n_{\rm jeans}$ was computed as a function of $m_{\rm DM}$ and the transfer cross-section $\sigma_T$ with rock in Ref.~\cite{Neufeld:2018slx}. There, it was noted that density enhancements $\eta_{\rm jeans} \equiv \frac{n_{\rm jeans}}{n_{\rm vir}}$ could be as large as $10^{14}$ for masses around a GeV, with severe loss due to sinking for masses $m_{\rm DM} \ge 10 \textrm{GeV}$. 

It was pointed out in Ref.~\cite{Pospelov:2019vuf}, that the depletion due to sinking is a very slow process at large cross-sections, i.e. the sinking terminal velocity $v_{\rm term} \ll v_{\rm th} \ll v_{\rm vir}$. As a result, for larger masses, there is a dark matter ``traffic jam'' on the Earth's crust, leading to an enhanced traffic jam density $\eta_{\rm tj}\equiv \frac{n_{\rm tj}}{n_{\rm vir}}=\frac{v_{\rm vir}}{v_{\rm term}}$. While this never reaches $\eta_{\rm jeans}$, it can still be large $\eta_{\rm tj} \approx \mathcal{O}\left(10^6\right)$ for large enough $\sigma_T$.

The local enhancements we use for the rest of this section is,

\begin{equation}
    \eta_{\rm local}\left(\sigma_T,m_{\rm DM}\right)=\eta_{\rm jeans}\left(\sigma_T,m_{\rm DM}\right)+\eta_{\rm tj}\left(\sigma_T,m_{\rm DM}\right)
    \label{eqn:nlocal}
\end{equation}

As a result, the locally thermalized DM has a mean-free-path $\lambda=\frac{1}{n_{\rm shield} \sigma_T} \gtrsim 30 \textrm{ cm} \gtrsim R_{\rm shield}$, where $R_{\rm shield}$, is the size of the cryogenic shielding. Hence, the DM with temperature $T_{\rm wall}$ arrives at the crystal detector without significant kinetic energy loss and can deposit that energy on the detector. The energy deposition rate for the general velocity dependent cross-section $\sigma_T=\sigma_0 v^n$, is given by~\cite{Dvorkin:2013cea},
\begin{equation}
\frac{dH}{dt} = \langle\sigma E v_{\rm rel}\rangle n_{\rm local} N_T = \frac{2^{\frac{n+5}{2}}\Gamma(3+\frac{n}{2})}{\sqrt{\pi}} n_{\rm local} N_T \frac{m_\chi m_T \sigma_0}{\left(m_
\chi+m_T\right)^2}\left(\frac{T_\chi}{m_\chi}+\frac{T_T}{m_T}\right)^\frac{n+1}{2}\left(T_\chi-T_T\right)
\label{heat0}
\end{equation}

Here, $v_{\rm rel}$ is the relative velocity between DM and target atoms, $n_T$ is the number of target atoms in the crystal, $m_T$ is the mass of the target atom. We will assume $n=0$ for contact interactions. The above expression was derived in Ref.~\cite{Dvorkin:2013cea} under the assumption that both gases are free. While obvious for the DM,  this assumptions for the crystal is not as straightforward. However, this is a good approximation as long as the typical momentum transfer resolves the individual atoms. In our case, 
\begin{equation}
    q_{\rm th}\approx 2\mu v_{\rm rel}=7~\textrm{keV} \sqrt{\frac{m}{\rm GeV}}
\end{equation}
is well above the $\sim$ 1 keV momentum transfers needed to resolve individual atoms in a crystal. 

Thus, setting n=0 in Eqn.~\ref{heat0}, the heating rate per target atom is given by,
\begin{align}
    \frac{dH}{dt}&=\frac{16}{\sqrt{2\pi}}\sigma_T n_{\rm local} \frac{\mu^2}{m_\chi m_T}\sqrt{\left(\frac{T_\chi}{m_\chi}+\frac{T_T}{m_T}\right)}\left(T_\chi-T_T\right) 
    \label{eqn:contactheat}
\end{align}
Here $T_\chi$, $T_T$ are temperatures of the dark matter and cryogenic crystal, $m_T$.

\begin{figure}
    \centering
     \includegraphics[width=0.78\textwidth]{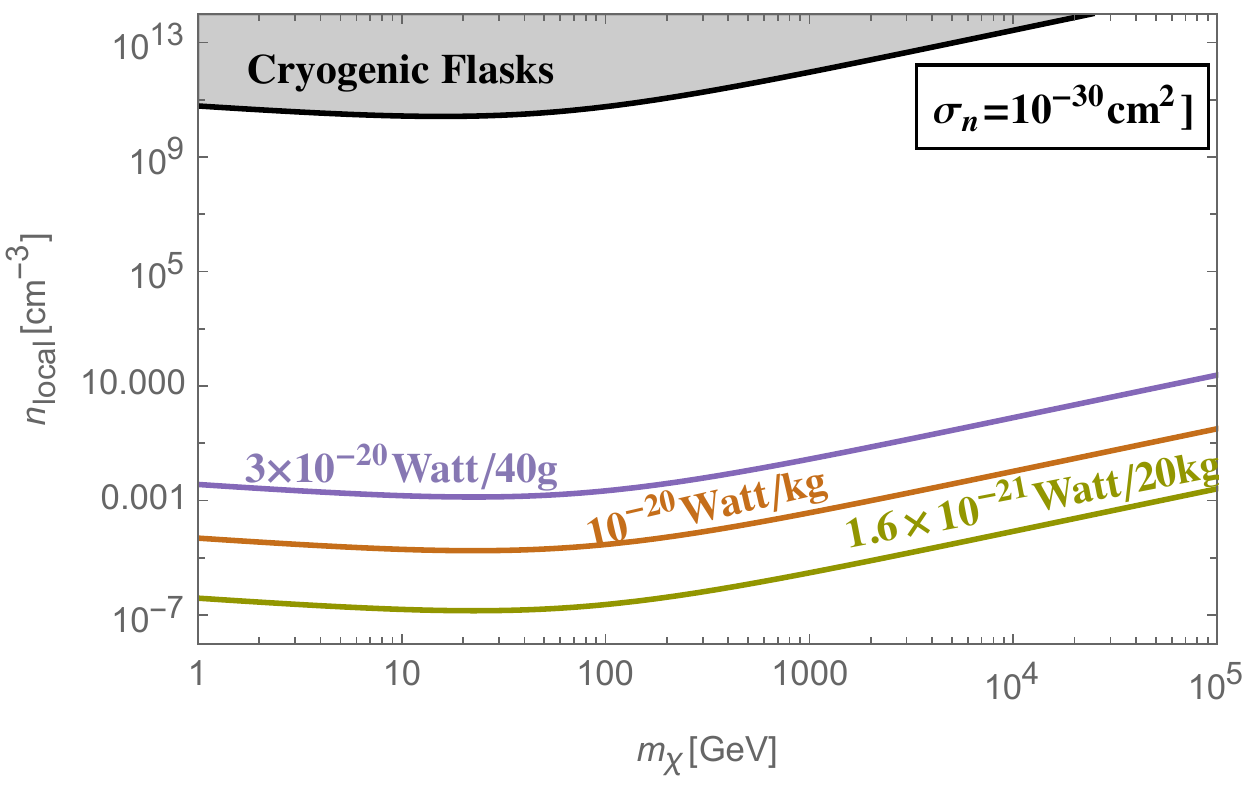}
     \caption{Projections for local thermal DM number densities $n_{\rm local}$ as a function of DM mass $m_\chi$ for various heat sensitivities are shown for a representative per-nucleon cross-section $\sigma_n=10^{-30}~\textrm{cm}^2$. Also shown for comparison is the existing limit from cryogenic flasks~\cite{Neufeld:2019xes}.}
     \label{fig:localdens1}
\end{figure}

\begin{figure}
     \includegraphics[width=0.78\textwidth]{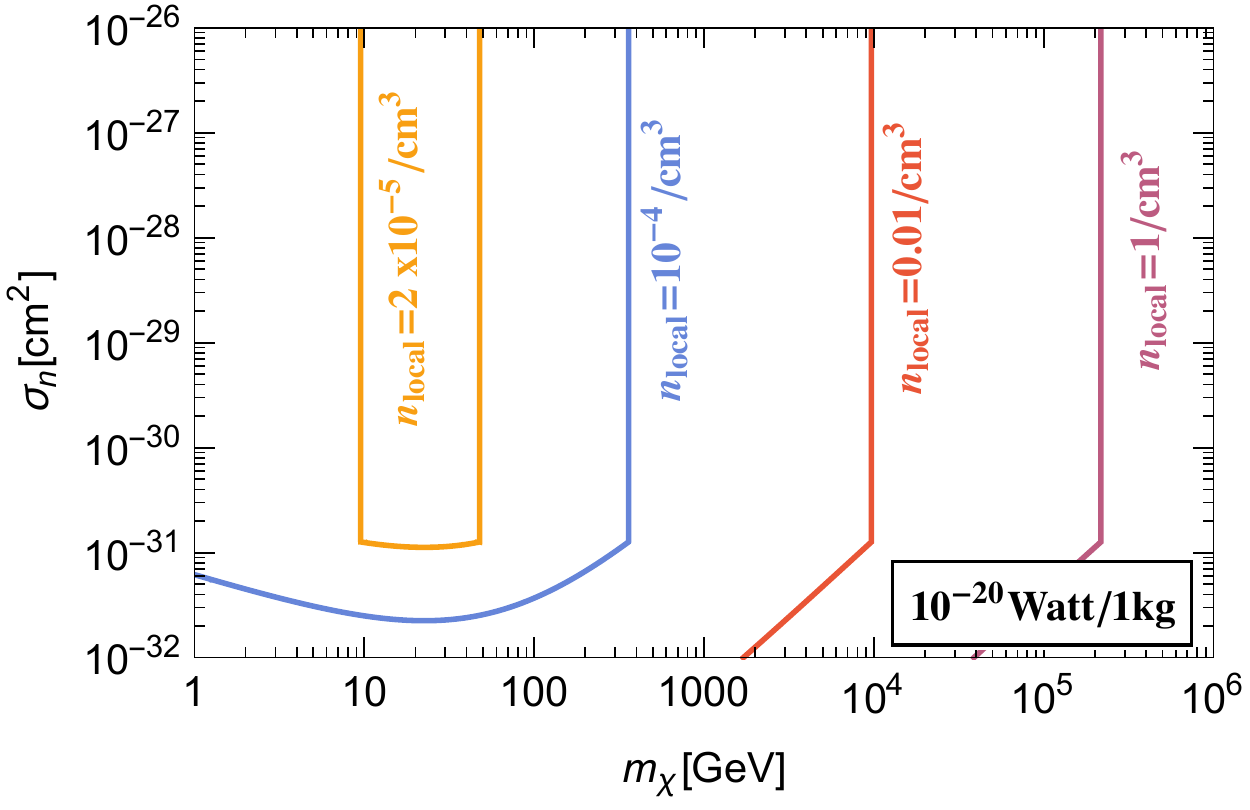}
    \caption{Projections for sensitivities to the local thermalized DM number density $n_{\rm local}$ are shown in the cross-section $\sigma_n$ vs DM mass $m_\chi$ plane for $10^{-20}~\textrm{Watt/1kg}$ heat sensitivity.}
       \label{fig:localdens2}
\end{figure}

Now for the saturated cross-section $\sigma_T\rightarrow 4\pi R_A^2$ and setting $T_\chi \gg T_T$, $m_\chi \gtrsim m_T$, we get,

\begin{align}
    \frac{dH}{dt}&= \frac{10^{-20}\textrm{Watt}}{1 \textrm{kg}}\left(\frac{A_T}{72}\right)^\frac{2}{3}\left(\frac{\rm TeV}{m_\chi}\right)^\frac{3}{2} \frac{n_{\rm local}}{10^{-3}~\textrm{cm}^{-3}}
\end{align}
Thus we see that with $10^{-20}\textrm{ Watt}$ sensitivities for a 1 kg sample, number densities larger than $n_{\rm local}=10^{-3}/\textrm{cm}^3$ can be probed at masses $m_\chi=1\textrm{ TeV}$. In Figs.~\ref{fig:localdens1} \&~\ref{fig:localdens2} we show the resultant limits for the three benchmark heating rates discussed in Section.~\ref{sec:technical}. To summarize, these are: $3\times10^{-20}~\textrm{Watt}/40\textrm{g}$ for an above-ground detector, $10^{-20}~\textrm{Watt}/\textrm{kg}$ for an underground detector and $1.86\times 10^{-21}~\textrm{Watt}/\textrm{20kg}$ for an idealized detector underground.

In Fig.~\ref{fig:localdens1}, we show the reach for these heat sensitivities to the local dark matter density $n_{\rm local}$ as a function of the dark matter mass $m_\chi$. The per-nucleon cross-section is assumed to be $\sigma_n=10^{-30}~\textrm{cm}^2$. These are compared with existing constraints from cryogenic flasks~\cite{Neufeld:2019xes} (gray). We see that even a modest experimental sensitivity from the above-ground  $3\times10^{-20}~\textrm{Watt}/40\textrm{g}$ benchmark can improve upon existing limits by about 8 orders of magnitude. Further improvements can lead to sensitivity to tiny number densities with the idealized detector sensitive to number densities as small as $n_{\rm local}=10^{-7}~\textrm{cm}^{-3}$.

In Fig.~\ref{fig:localdens2}, contours of $n_{\rm local}$, the dark matter density in the lab are plotted in the cross-section $\sigma_n$ vs DM mass $m_\chi$ parameter space for $10^{-20}~\textrm{Watt/1kg}$ sensitivity. The saturation of the cross-section which is capped at the geometric size of the nucleus causes the flat behavior above $\sigma_n \gtrsim 10^{-31}~\textrm{cm}^2$. The peak sensitivity is obtained when $m_\chi\approx m_T$ where there is kinematic matching. 

We next use the enhancements in Eqn.~\ref{eqn:nlocal} to project limits on the ambient virial DM fraction $f_{\rm DM}=\frac{\rho_
\chi}{\rho_{\rm DM}}$. The above ground benchmark is assumed to have 1 meter shielding while the underground benchmarks are assumed to have 1 km shielding for the estimation of traffic-jam densities. 

\begin{figure}
    \centering
    \includegraphics{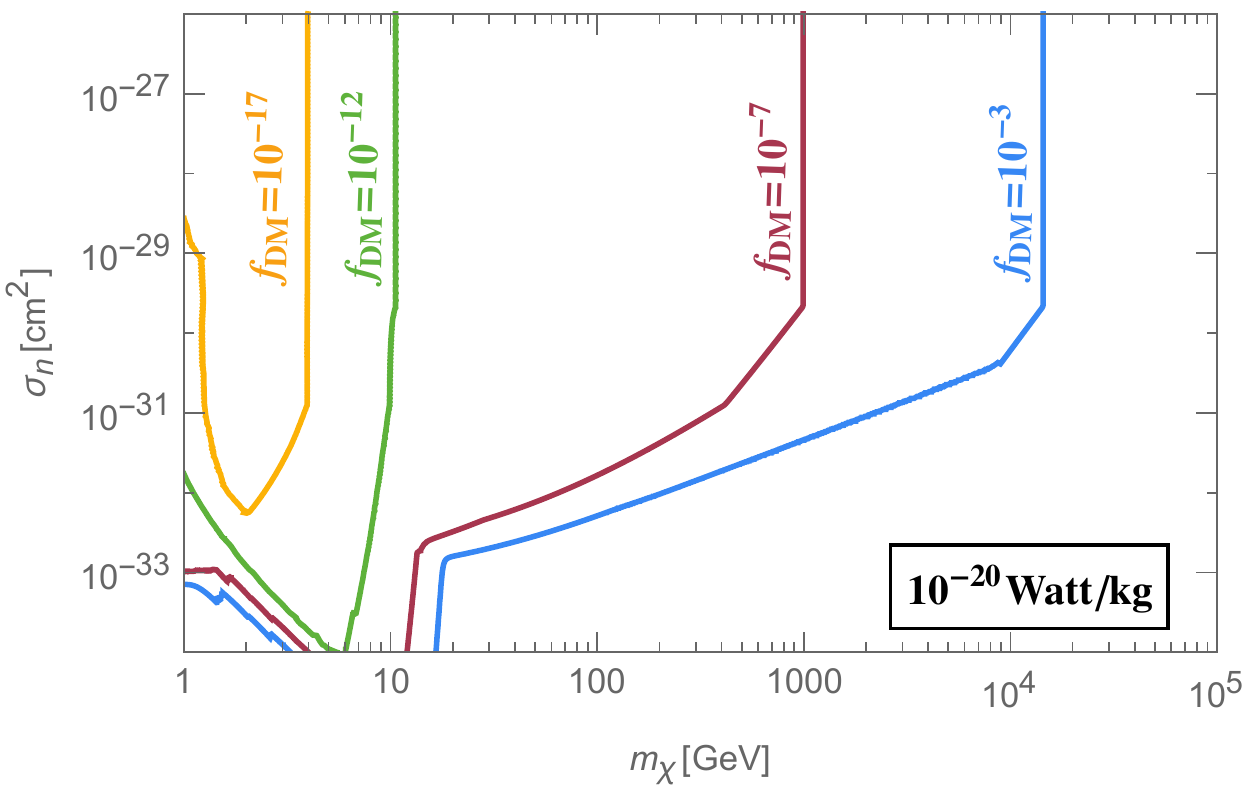}
    \caption{Projections are shown for the DM fraction $f_{\rm DM}\equiv \frac{\rho_\chi}{\rho_{\rm DM}}$  in the per-nucleon cross-section $\sigma_n$ vs DM mass $m_\chi$ plane for $10^{-20}~\textrm{Watt/1kg}$ heat sensitivity.}
    \label{fig:fdm}
\end{figure}

In Fig.~\ref{fig:fdm} we plot contours of $f_{\rm DM}\equiv \frac{\rho_\chi}{\rho_{\rm DM}}$ that can be probed by the $10^{-20}~\textrm{Watt}/\textrm{1kg}$ underground heat sensitivity benchmark. We see that fractions as small as $f_{\rm DM}\approx 10^{-17}$ can be probed in the $1-6~\textrm{GeV}$ window. Fractions as small as $f_{\rm DM}\approx10^{-7}$ can be probed upto $m_\chi=2~\textrm{TeV}$.

\begin{figure}
    \centering
   \begin{subfigure}{0.78\textwidth}
   \includegraphics[width=0.98\textwidth]{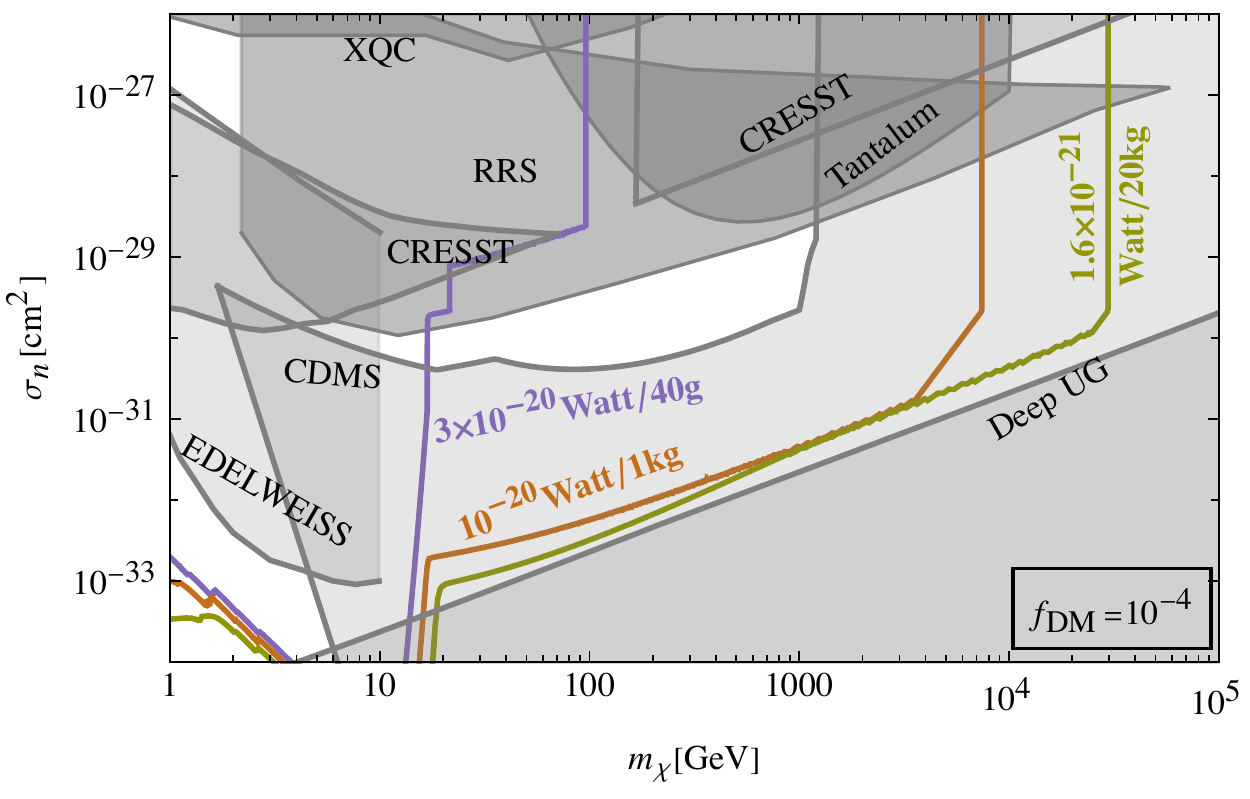}
   \end{subfigure}
   \begin{subfigure}{0.78\textwidth}
  \includegraphics[width=0.98\textwidth]{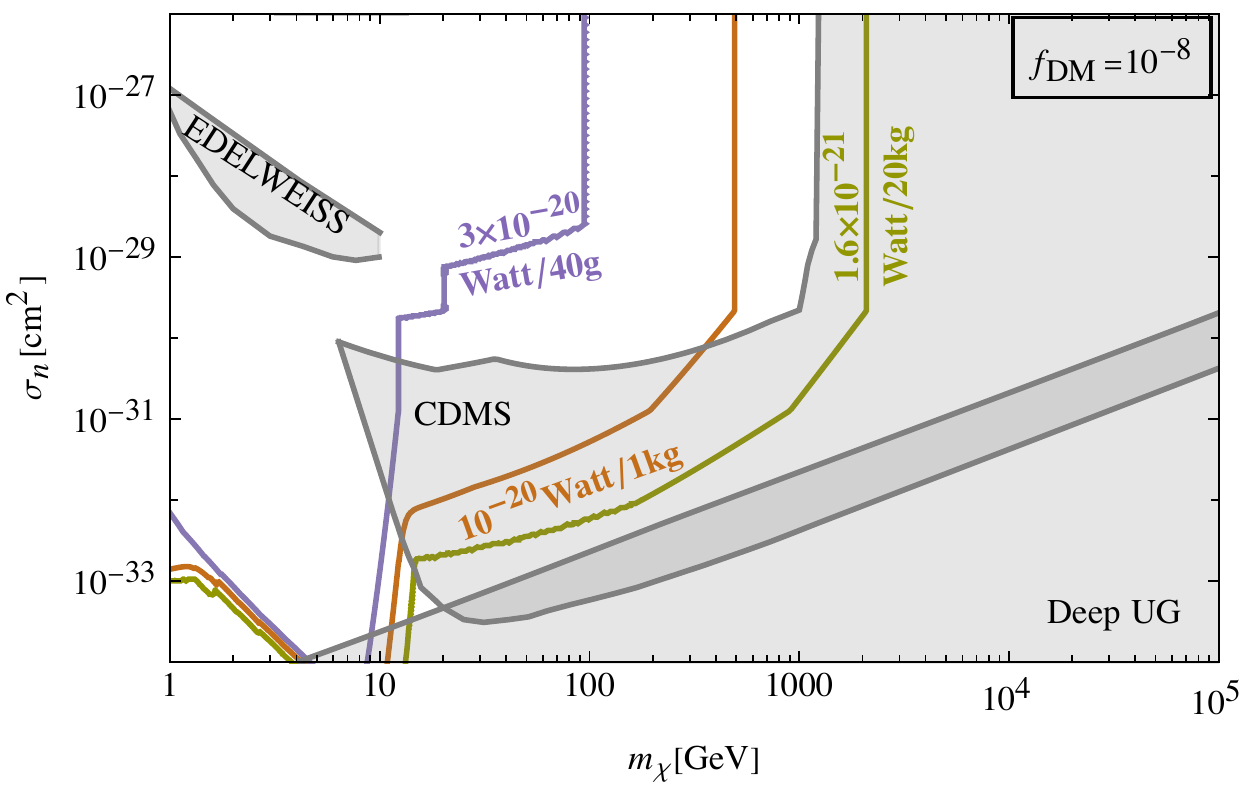}
   \end{subfigure}
 \caption{Projections are shown for two different DM fractions $f_{\rm DM}\equiv \frac{\rho_\chi}{\rho_{\rm DM}}$, $f_{\rm DM}=10^{-4}$~(\textbf{Top}) and $f_{\rm DM}=10^{-8}$~(\textbf{Bottom}). The axes are the per-nucleon cross-section $\sigma_n$ vs DM mass $m_\chi$. Contours of $3\times 10^{-20}~\textrm{Watt/40g}$ (Purple), $10^{-20}~\textrm{Watt/1kg}$ (Orange) and $1.6\times10^{-21}~\textrm{Watt/20kg}$ (Green) heat sensitivity are shown. Also shown are existing limits from  the rocket experiment (XQC)~\cite{mccammon2002high}, balloon experiment (RRS)~\cite{rich1987search}, isomeric Tantallum~\cite{Lehnert:2019tuw}, Surface Experiments (CRESST~\cite{CRESST:2017ues} and CDMS~\cite{CDMS:2002moo}) and Deep UG experiments.}
     \label{fig:fdm2}
   \end{figure}

We compare the the benchmark heat sensitivies for constant DM fractions $f_{\rm DM}=10^{-4}~\left(10^{-8}\right)$ in Fig.~\ref{fig:fdm2} Top (Bottom) to existing limits from the rocket experiment (XQC)~\cite{mccammon2002high}, balloon experiment (RRS)~\cite{rich1987search}, isomeric Tantallum~\cite{Lehnert:2019tuw}, Surface Experiments (CRESST~\cite{CRESST:2017ues}, EDELWEISS~\cite{EDELWEISS:2019vjv} and CDMS~\cite{CDMS:2002moo}) and Deep UG experiments. As seen in Fig.~\ref{fig:fdm2}~(Top), for $f_{\rm DM}=10^{-4}$, the open window for DM that strongly interacts with the SM can be probed upto masses as large as $m_\chi=10~\textrm{TeV}$ with heat sensitivity $10^{-20}~\textrm{Watt/1kg}$. While existing limits largely disappear for DM fractions below $f_{\rm DM}=10^{-8}$ as seen in Fig.~\ref{fig:fdm2}~(Bottom), this region will be probed even with the above the ground $3\times10^{-20}~\textrm{Watt/40g}$ sensitivity. 

It is important to point out that the traffic jam effect which determines $n_{\rm local}$ at masses above $5~\textrm{GeV}$ is depth dependent with increase in density at lower depths for the same cross-section and mass. For e.g. for $m_\chi=100$ GeV and $\sigma_n=10^{-30}\textrm{cm}^2$, the enhancement at the surface is very close to unity, whereas at the depth of 1km, it is large, $\eta\approx7.5\times10^6$. Hence, an underground experiment, apart from its background amelioration capabilities also enjoys a superior local dark matter number density it can uncover. 


\subsection{Dark Photon}
While the primary motivation of our calorimetric detector is to probe large DM-SM interactions, this kind of detector can also probe other well motivated DM candidates that have a large local abundance. As an example, consider light dark photons in the meV mass range that are kinetically mixed with the photon.  The relevant physics is captured in the Lagrangian,
\begin{align}
    \mathcal{L} \supset -\frac{\kappa}{2}F^{\mu\nu} F_{\mu \nu}'+\frac{1}{2}m_{A'}^2 A'^2
\end{align}
here $\kappa$ is the kinetic mixing parameter, $m_{A'}$ is the mass of the dark photon $A'$ and $F$ and $F'$ are the field strength tensors of the SM and dark photons. 

The absorption cross-section $\sigma_{A'}$ for dark photons in a material can be easily estimated from data on SM photon cross-section $\sigma_\gamma$ using the relation~\cite{Griffin:2020lgd},
\begin{equation}
    \sigma_{A'} v_{\rm vir}=\frac{\kappa^2}{\tilde{\epsilon}^2} \sigma_{\gamma}
\end{equation}
here,  $v_{\rm vir}$ is the virial DM velocity and $\tilde{\epsilon}$ is the dielectric function that captures in-medium effects. 
Thus, the total heating rate is given by,
\begin{equation}
    \frac{dH}{dt}=\frac{\kappa^2}{\tilde{\epsilon}^2} \rho_{\rm DM}  \sigma_{\gamma}
    \label{dpeqn}
\end{equation}
\begin{figure}
    \centering
    \includegraphics{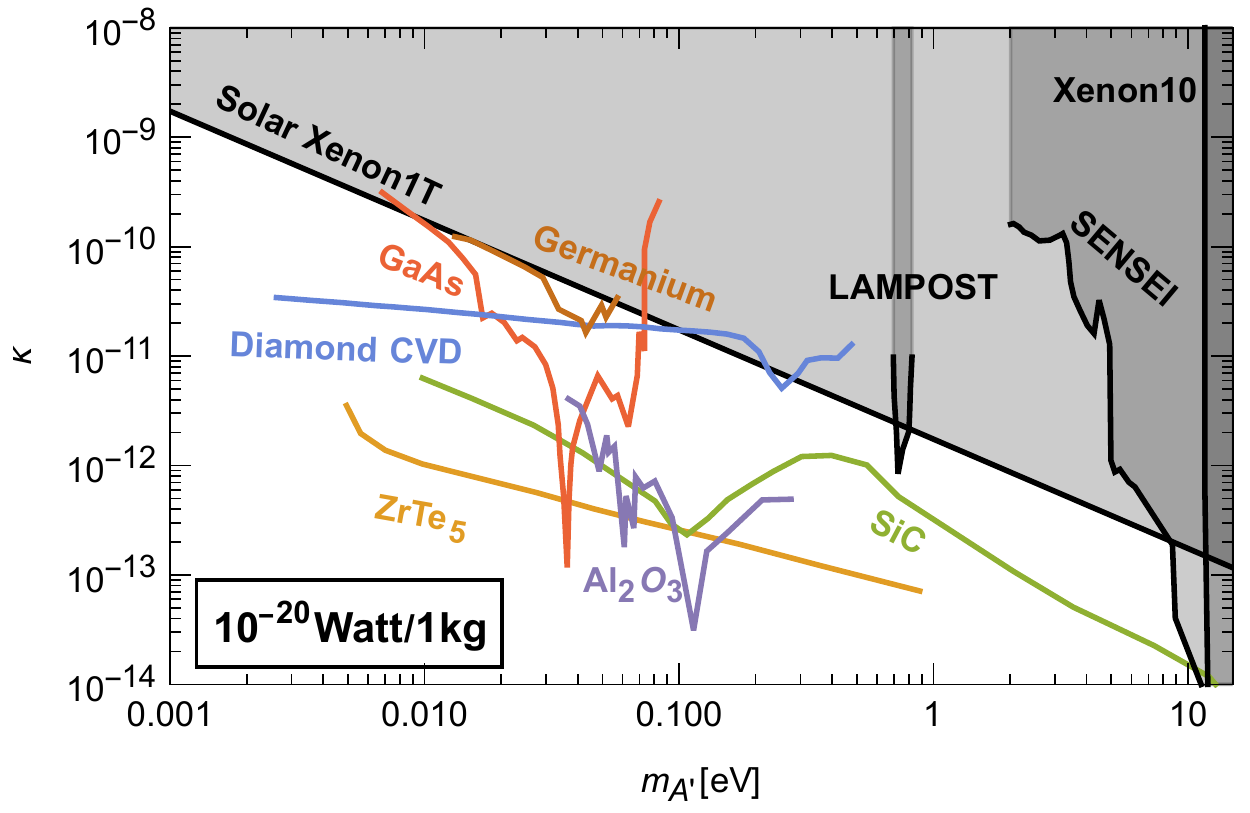}
    \caption{Projections for reach in the kinetic mixing parameter $\kappa$ are shown as a function of the dark photon mass $m_{A'}$ for heat sensitivity $10^{-20}~\textrm{Watt/1kg}$ for different materials. Absorption data is obtained from Sapphire($\textrm{Al}_2\textrm{O}_3$)~\cite{Griffin:2018bjn}, Gallium Arsenide~\cite{Griffin:2018bjn}, $\textrm{ZrTe}_5$~\cite{Chen:2022pyd}, Silicon Carbide~\cite{Griffin:2020lgd}, Diamond Chemical Vapor Deposition (CVD)~\cite{Kurinsky:2019pgb} and Germanium~\cite{Hochberg:2016sqx}. }
    \label{fig:darkphoton}
\end{figure}
Many candidate materials have been analyzed in literature in the context of dark photon absorption using single phonon rates.  These include Sapphire($\textrm{Al}_2\textrm{O}_3$)~\cite{Griffin:2018bjn}, Gallium Arsenide~\cite{Griffin:2018bjn}, $\textrm{ZrTe}_5$~\cite{Chen:2022pyd}, Silicon Carbide~\cite{Griffin:2020lgd}, Diamond Chemical Vapor Deposition (CVD)~\cite{Kurinsky:2019pgb} and Germanium~\cite{Hochberg:2016sqx}. We recast these analyses into projections for the $10^{-20}~\textrm{Watt/1 kg}$ setup in Fig.~\ref{fig:darkphoton}, where we see that calorimetric detectors are able to probe dark photon dark matter models well beyond present limits. Recasting to other heat sensitivity benchmarks is straightforward with $\kappa\propto \left(\frac{dH}{dt}\right)^\frac{1}{2}$ as seen in Eqn.~\ref{dpeqn}. 

\section{Conclusions}
\label{sec:conclusions}

The methods described in this paper are well suited to probe dark matter particles whose energy depositions are too soft to cause ionization or scintillation, but whose interaction rate with the standard model is high enough to deposit detectable amounts of heat in a suitably optimized detector. As such, this detection concept lies between conventional WIMP dark matter detectors where the individual collisions of dark matter are hard enough to cause ionization/scintillation and detection methods for ultra-light dark matter where the softness and rarity of individual scattering events necessitates the search for collective effects of the dark matter on materials. 

The sensitivity estimates we have made in this paper are based on experience and measurement with low thresholds calorimetric detectors such as CRESST, EDELWEISS and SuperCDMS. Our signal lacks distinguishing features such as ionization and scintillaton. Thus, to reliably probe this signal, a dedicated apparatus is necessary where we are able to compare the heat in the thermal sensors with the target crystal in and out of contact with the thermal sensors. Importantly, even though this requires a dedicated setup, the technology  necessary for the proposed experiment is within the technical scope of the technology envisioned for CRESST, EDELWEISS and SuperCDMS. 

There are natural classes of dark matter that can be probed by such a detector. In this paper, we have made estimates of the reach of this class of detector in probing strongly interacting dark matter and ultra-light hidden photon dark matter. Other classes of dark matter that could potentially be detected using this concept includes composite dark matter blobs, mini-clusters and milli-charged particles~\cite{Budker:2021quh}. In future work, we intend to explicitly compute the reach of this kind of detector concept in probing these models. This detector may also serve as a stringent test of solutions to the neutron bottle anomaly that involve the scattering of dark matter \cite{Rajendran:2020tmw}.  This detection concept may also be useful in searching for dark radiation (for example, from dark energy \cite{Berghaus:2020ekh} or from the early universe) since these relativistic degrees of freedom will have short coherence times and are thus difficult to detect using conventional detection methods for ultra-light dark matter that typically leverage the long coherence time of the signal to build a detectable response in a detector. 

\section*{Acknowledgments}
S.R.~ is supported in part by the U.S.~National Science Foundation (NSF) under Grant No.~PHY-1818899. This work was supported by the U.S.~Department of Energy (DOE), Office of Science, National Quantum Information Science Research Centers, Superconducting Quantum Materials and Systems Center (SQMS) under contract No.~DE-AC02-07CH11359. 
S.R.~is also supported by the DOE under a QuantISED grant for MAGIS, and the Simons Investigator Award No.~827042. H.R.~acknowledges the support from the Simons Investigator Award 824870, DOE Grant DE-SC0012012, NSF Grant PHY2014215, DOE HEP QuantISED award no. 100495,
and the Gordon and Betty Moore Foundation Grant
GBMF7946.

\bibliography{thermal}

\appendix

\section{Detector thermal model parameters}

We hereafter list the thermal model parameters considered in Sec.~\ref{sec:technical} with the Figures~\ref{fig:NEP_IdealDetector} and \ref{fig:NEP_RealDetectors}. Parameter values regarding the NTD electron-phonon and glue thermal couplings are taken from~\cite{novati:tel-01963790} and references therein. Other relevant thermal conductivities follow the measurements from~\cite{marnieros:tel-01088881,Pinckney:2021wtd}. The NTD $R_0$ and $T_0$ parameters are from~\cite{Billard:2016apk} and are consistent with those presented in~\cite{novati:tel-01963790, Mathimalar:2014sfa}. Eventually, the voltage and current noise from the preamplifier considering a 200~pF HEMT are taken from~\cite{RICOCHET:2021fix}.

\setlength{\tabcolsep}{0.1cm}
\renewcommand{\arraystretch}{1.}
\begin{table}
\begin{center}
\hspace*{-0.5cm}
\begin{tabular}{c|c|c}
\hline
 Component  &  Value & Notes\\ \hline \hline  
{\it \bf Bath} & & \\ 
Temperature & 10 mK &  \\ \hline
{\it \bf Absorber} & Ge  & \\
Volume & $\pi\times 16^2/4\times 19 = 3820$ cm$^3$ & 20~kg \\ 
Heat capacity & $C_a = 1.18\times10^{-8}$ J/K & $C_a = 2.7\times 10^{-6} \bar{T}_a^3$ J/K/cm$^3$  \\ 
Surface gold pads & $2 \times (40 \times 40) = 3200$~mm$^2$ & Connected to sapphire slabs\\ \hline
{\it \bf Sapphire slabs} & Al$_{2}$O$_{3}$  & \\
Volume & $10 \times 2.5 \times 0.1 = 2.5$ cm$^3$ & 9.95~g \\ 
Heat capacity & $C_{s_{i}} = 1.02\times10^{-12}$ J/K & $C_s = 0.35\times 10^{-6} \bar{T}_{s_{i}}^3$ J/K/cm$^3$  \\ 
Surface ($S_{AuGe_i}$) & $10 \times 40 = 400$~mm$^2$ & Connected to Ge crystal \\
Surface ($S_{AuB_i}$) & $25 \times 25 = 625~\mu m^2$ & Connected to bath \\ \hline
{\it \bf NTD} & Ge & \\
Surface ($S_{NTD}$) & $20\times20 = 400$ mm$^2$ & \\
Volume ($V_{NTD}$) & $S_{NTD} \times 1 = 400$ mm$^3$ &  \\ 
$R_0$ / $T_0$ & 0.96 $\Omega$ / 4.52 K & $\bar{R}(\bar{T}_{e_{i,j}}) = R_0e^{\sqrt{T_0/\bar{T}_{e_{i,j}}}}$ \\ 
Heat capacity (phonon) & $C_p = 1.26\times10^{-12}$ J/K & $C_p = 2.7\times 10^{-6} \bar{T}_p^3$ J/K/cm$^3$  \\
Heat capacity (electron) & $C_e = 4.63\times10^{-9}$ J/K & $C_e = 1.1\times 10^{-6} \bar{T}_e$ J/K/cm$^3$  \\  \hline
{\it \bf Conductivities} & & $G = dP/dT = n\times g \times T^{n-1}\times S {\rm(or~V)}$\\
Electron-Phonon (NTD) & $G_{ep} = 3.12\times 10^{-8}$~W/K & $g_{ep}$ = 100 W/K$^6$/cm$^3$  \\
Glue (NTD-Sapp.)  & $G_{ps} = 2.23\times 10^{-6}$~W/K &  $g_{glue} = 1.4\times 10^{-4} $ W/K$^{3.5}$/mm$^2$ \\
Au pads (Sapp.-Ge)  & $G_{AuGe} = 2.34\times 10^{-7}$~W/K &  $g_k = 1.25\times10^{-4}$ W/K$^4$/mm$^2$ \\
Au pads (Sapp.-Bath)  & $G_{AuB} = 3.65 \times 10^{-13}$~W/K &  $g_k = 1.25\times10^{-4}$ W/K$^4$/mm$^2$ \\
Sapphire balls (Ge-Bath)  & $G_{GeB} = 5.60 \times 10^{-12}$ W/K &  $g_{b} = 1.33\times10^{-7}$ W/K$^4$/ball ($N_{GeB} = 9$)  \\  \hline
{\it \bf Equilibrium state} &   & \\
NTD-electron & $\bar{T}_e$ = 10.54 mK &  \\
NTD-phonon & $\bar{T}_p$ = 10.54 mK &  \\
Sapphire slabs & $\bar{T}_s$ = 10.54 mK &  \\ 
Absorber & $\bar{T}_a$ = 10.54 mK &  \\ 
Voltage & $\bar{V}$ = 0.86 mV &  $\bar{V} = V_b\bar{R}/(R_L + \bar{R})$\\ \hline
{\it \bf Electronic considerations} &   & \\
Voltage bias & $V_b$ = 10~mV & $I_p = V_b/(R_L + \bar{R}) \approx 0.91$ pA    \\
Load resistor & $R_L$ = 10 G$\Omega$ & $T_{R_L}$ = 10 mK   \\
NTD Resistance & $\bar{R}$ = 947 M$\Omega$ &  $\bar{R} = R(\bar{T}_e)$\\
NTD Joule power & $P_J$ = 0.79~fW & $P_J = \bar{R}I_p^2$   \\
200pF-HEMT ($C_{c_{i,j}}$ = 250 pF) & $e^2_n = e^2_a + e^2_b/f $ & $\{e_a, e_b\} = \{0.18, 5.2\}$  nV/$\sqrt{\rm{Hz}}$    \\
 & $i^2_n = i^2_a + i^2_b\times f $ & $\{i_a, i_b\} = \{8.2\times 10^{-4}, 21\}$  aA/$\sqrt{\rm{Hz}}$    \\
 \hline
{\it \bf Time constants of the system} &   & driven by :\\
Rise time & $\tau_r = 225$ ms  &  $\tau_r = \bar{R}\times C_c \approx 246$~ms \\
Decay time & $\tau_d = 3417$~s  &    $\tau_{d} = C_{\rm tot}/(G_{AuB_a}\parallel G_{AuB_b} \parallel G_{GeB}) \approx 3760$~s  \\ \hline

{\it \bf Detector performance} &    &   \\
NEP($\omega=0$) & $2.9\times 10^{-19}$~W/$\sqrt{\rm Hz}$  &  Assuming AC modulation to cancel $e_n$\\
DC power sensitivity (5-$\sigma$) &  $1.86\times 10^{-21}$~W/(20 kg) &  After 1 week of DM search \\
Energy resolution & $\sigma_E = 101$~eV  &  Energy deposition in target absorber\\
Timing resolution & $\sigma_t = 2259$~s.eV  &  $\sigma_t \approx 2.2$~s at 1~keV\\
\hline
\end{tabular}
\caption{Characteristics of the thermal model simulation shown in Fig.~\ref{fig:NEP_IdealDetector}}
\label{tab:parameter_IdealDetector}
\end{center}
\end{table}

\setlength{\tabcolsep}{0.1cm}
\renewcommand{\arraystretch}{1.}
\begin{table}
\begin{center}
\hspace*{-0.5cm}
\begin{tabular}{c|c|c}
\hline
 Component  &  Value & Notes\\ \hline \hline  
{\it \bf Bath} & & \\ 
Temperature & 10 mK &  \\ \hline
{\it \bf Absorber} & Ge  & \\
Volume & $\pi\times 3^8/4\times 4 = 192$~cm$^3$ & 1~kg \\ 
Heat capacity & $C_a = 6.00\times10^{-10}$ J/K & $C_a = 2.7\times 10^{-6} \bar{T}_a^3$ J/K/cm$^3$  \\ 
Surface gold pads & $2 \times (10 \times 10) = 200$~mm$^2$ & Connected to sapphire slabs\\ \hline
{\it \bf Sapphire slabs} & Al$_{2}$O$_{3}$  & \\
Volume & $10 \times 2.5 \times 0.1 = 2.5$ cm$^3$ & 9.95~g \\ 
Heat capacity & $C_{s_{i}} = 1.01\times10^{-12}$ J/K & $C_s = 0.35\times 10^{-6} \bar{T}_{s_{i}}^3$ J/K/cm$^3$  \\ 
Surface ($S_{AuGe_i}$) & $10 \times 10 = 100$~mm$^2$ & Connected to Ge crystal \\
Surface ($S_{AuB_i}$) & $0.3 \times 0.5 = 0.3$~mm$^2$ & Connected to bath \\ \hline
{\it \bf NTD} & Ge & \\
Surface ($S_{NTD}$) & $5\times5 = 25$ mm$^2$ & \\
Volume ($V_{NTD}$) & $S_{NTD} \times 1 = 25$ mm$^3$ &  \\ 
$R_0$ / $T_0$ & 0.96 $\Omega$ / 4.52 K & $\bar{R}(\bar{T}_{e_{i,j}}) = R_0e^{\sqrt{T_0/\bar{T}_{e_{i,j}}}}$ \\ 
Heat capacity (phonon) & $C_p = 7.8\times10^{-14}$ J/K & $C_p = 2.7\times 10^{-6} \bar{T}_p^3$ J/K/cm$^3$  \\
Heat capacity (electron) & $C_e = 2.89\times10^{-10}$ J/K & $C_e = 1.1\times 10^{-6} \bar{T}_e$ J/K/cm$^3$  \\  \hline
{\it \bf Conductivities} & & $G = dP/dT = n\times g \times T^{n-1}\times S {\rm(or~V)}$\\
Electron-Phonon (NTD) & $G_{ep} = 1.92\times 10^{-9}$~W/K & $g_{ep}$ = 100 W/K$^6$/cm$^3$  \\
Glue (NTD-Sapp.)  & $G_{ps} = 1.38\times 10^{-7}$~W/K &  $g_{glue} = 1.4\times 10^{-4} $ W/K$^{3.5}$/mm$^2$ \\
Au pads (Sapp.-Ge)  & $G_{AuGe} = 5.78\times 10^{-8}$~W/K &  $g_k = 1.25\times10^{-4}$ W/K$^4$/mm$^2$ \\
Au pads (Sapp.-Bath)  & $G_{AuB} = 8.67 \times 10^{-11}$~W/K &  $g_k = 1.25\times10^{-4}$ W/K$^4$/mm$^2$ \\
Sapphire balls (Ge-Bath)  & $G_{GeB} = 1.84 \times 10^{-12}$ W/K &  $g_{b} = 1.33\times10^{-7}$ W/K$^4$/ball ($N_{GeB} = 3$)  \\  \hline
{\it \bf Equilibrium state} &   & \\
NTD-electron & $\bar{T}_e$ = 10.50 mK &  \\
NTD-phonon & $\bar{T}_p$ = 10.49 mK &  \\
Sapphire slabs & $\bar{T}_s$ = 10.49 mK &  \\ 
Absorber & $\bar{T}_a$ = 10.49 mK &  \\ 
Voltage & $\bar{V}$ = 4.45 mV &  $\bar{V} = V_b\bar{R}/(R_L + \bar{R})$\\ \hline
{\it \bf Electronic considerations} &   & \\
Voltage bias & $V_b$ = 50~mV & $I_p = V_b/(R_L + \bar{R}) \approx 4.55$ pA    \\
Load resistor & $R_L$ = 10 G$\Omega$ & $T_{R_L}$ = 10 mK   \\
NTD Resistance & $\bar{R}$ = 977~M$\Omega$ &  $\bar{R} = R(\bar{T}_e)$\\
NTD Joule power & $P_J$ = 20.3~fW & $P_J = \bar{R}I_p^2$   \\
200pF-HEMT ($C_{c_{i,j}}$ = 250 pF) & $e^2_n = e^2_a + e^2_b/f $ & $\{e_a, e_b\} = \{0.18, 5.2\}$  nV/$\sqrt{\rm{Hz}}$    \\
 & $i^2_n = i^2_a + i^2_b\times f $ & $\{i_a, i_b\} = \{8.2\times 10^{-4}, 21\}$  aA/$\sqrt{\rm{Hz}}$    \\
 \hline
{\it \bf Time constants of the system} &   & driven by :\\
Rise time & $\tau_r = 161$~ms  &  $\tau_r = \bar{R}\times C_c \approx 254$~ms \\
Decay time & $\tau_d = 7.16$~s  &    $\tau_{d} = C_{\rm tot}/(G_{AuB_a}\parallel G_{AuB_b} \parallel G_{GeB}) \approx 6.92$~s  \\ \hline
{\it \bf Detector performance} &    &   \\
NEP($\omega=0$) & $1.62 \times 10^{-18}$~W/$\sqrt{\rm Hz}$  &  Assuming AC modulation to cancel $e_n$\\
DC power sensitivity (5-$\sigma$) &  $1.04\times 10^{-20}$~W/(1 kg) &  After 2 weeks of DM search (50\% livetime) \\
Energy resolution & $\sigma_E = 26.0$~eV  &  Energy deposition in target absorber\\
Timing resolution & $\sigma_t = 24.4$~s.eV  &  $\sigma_t \approx 24.4$ ~ms at 1~keV\\
\hline
\end{tabular}
\caption{Characteristics of the thermal model simulation shown in Fig.~\ref{fig:NEP_RealDetectors} (left panel)}
\label{tab:parameter_1kgDetector}
\end{center}
\end{table}

\setlength{\tabcolsep}{0.1cm}
\renewcommand{\arraystretch}{1.}
\begin{table}
\begin{center}
\hspace*{-0.5cm}
\begin{tabular}{c|c|c}
\hline
 Component  &  Value & Notes\\ \hline \hline  
{\it \bf Bath} & & \\ 
Temperature & 10 mK &  \\ \hline
{\it \bf Absorber} & Ge  & \\
Volume & $\pi\times 3^2/4\times 1 = 7 cm^3$ & 40~g \\ 
Heat capacity & $C_a = 3.82\times10^{-11}$ J/K & $C_a = 2.7\times 10^{-6} \bar{T}_a^3$ J/K/cm$^3$  \\ 
Surface gold pads & $2 \times (10 \times 10) = 200$~mm$^2$ & Connected to sapphire slabs\\ \hline
{\it \bf Sapphire slabs} & Al$_{2}$O$_{3}$  & \\
Volume & $10 \times 2.5 \times 0.1 = 2.5$ cm$^3$ & 9.95~g \\ 
Heat capacity & $C_{s_{i}} = 1.76\times10^{-12}$ J/K & $C_s = 0.35\times 10^{-6} \bar{T}_{s_{i}}^3$ J/K/cm$^3$  \\ 
Surface ($S_{AuGe_i}$) & $10 \times 10 = 100$~mm$^2$ & Connected to Ge crystal \\
Surface ($S_{AuB_i}$) & $0.6 \times 0.5 = 0.3~mm^2$ & Connected to bath \\ \hline
{\it \bf NTD} & Ge & \\
Surface ($S_{NTD}$) & $2\times2 = 4$ mm$^2$ & \\
Volume ($V_{NTD}$) & $S_{NTD} \times 1 = 4$ mm$^3$ &  \\ 
$R_0$ / $T_0$ & 0.96 $\Omega$ / 4.52 K & $\bar{R}(\bar{T}_{e_{i,j}}) = R_0e^{\sqrt{T_0/\bar{T}_{e_{i,j}}}}$ \\ 
Heat capacity (phonon) & $C_p = 2.18\times10^{-14}$ J/K & $C_p = 2.7\times 10^{-6} \bar{T}_p^3$ J/K/cm$^3$  \\
Heat capacity (electron) & $C_e = 5.72\times10^{-11}$ J/K & $C_e = 1.1\times 10^{-6} \bar{T}_e$ J/K/cm$^3$  \\  \hline
{\it \bf Conductivities} & & $G = dP/dT = n\times g \times T^{n-1}\times S {\rm(or~V)}$\\
Electron-Phonon (NTD) & $G_{ep} = 8.92\times 10^{-10}$~W/K & $g_{ep}$ = 100 W/K$^6$/cm$^3$  \\
Glue (NTD-Sapp.)  & $G_{ps} = 3.53\times 10^{-8}$~W/K &  $g_{glue} = 1.4\times 10^{-4} $ W/K$^{3.5}$/mm$^2$ \\
Au pads (Sapp.-Ge)  & $G_{AuGe} = 1.01\times 10^{-7}$~W/K &  $g_k = 1.25\times10^{-4}$ W/K$^4$/mm$^2$ \\
Au pads (Sapp.-Bath)  & $G_{AuB} = 3.03 \times 10^{-10}$~W/K &  $g_k = 1.25\times10^{-4}$ W/K$^4$/mm$^2$ \\
Sapphire balls (Ge-Bath)  & $G_{GeB} = 3.22 \times 10^{-12}$ W/K &  $g_{b} = 1.33\times10^{-7}$ W/K$^4$/ball ($N_{GeB} = 3$)  \\  \hline
{\it \bf Equilibrium state} &   & \\
NTD-electron & $\bar{T}_e$ = 13.00 mK &  \\
NTD-phonon & $\bar{T}_p$ = 12.65 mK &  \\
Sapphire slabs & $\bar{T}_s$ = 12.64 mK &  \\ 
Absorber & $\bar{T}_a$ = 12.64 mK &  \\ 
Voltage & $\bar{V}$ = 5.93 mV &  $\bar{V} = V_b\bar{R}/(R_L + \bar{R})$\\ \hline
{\it \bf Electronic considerations} &   & \\
Voltage bias & $V_b$ = 500~mV & $I_p = V_b/(R_L + \bar{R}) \approx 49.4$ pA    \\
Load resistor & $R_L$ = 10 G$\Omega$ & $T_{R_L}$ = 10 mK   \\
NTD Resistance & $\bar{R}$ = 120~M$\Omega$ &  $\bar{R} = R(\bar{T}_e)$\\
NTD Joule power & $P_J$ = 293~fW & $P_J = \bar{R}I_p^2$   \\
200pF-HEMT ($C_{c_{i,j}}$ = 250 pF) & $e^2_n = e^2_a + e^2_b/f $ & $\{e_a, e_b\} = \{0.18, 5.2\}$  nV/$\sqrt{\rm{Hz}}$    \\
 & $i^2_n = i^2_a + i^2_b\times f $ & $\{i_a, i_b\} = \{8.2\times 10^{-4}, 21\}$  aA/$\sqrt{\rm{Hz}}$    \\
 \hline
{\it \bf Time constants of the system} &   & driven by :\\
Rise time & $\tau_r = 36.2$ ms  &  $\tau_r = \bar{R}\times C_c \approx 31.2$~ms \\
Decay time & $\tau_d = 142.2$~ms  &    $\tau_{d} = C_{\rm tot}/(G_{AuB_a}\parallel G_{AuB_b} \parallel G_{GeB}) \approx 230$~ms  \\ \hline

{\it \bf Detector performance} &    &   \\
NEP($\omega=0$) & $4.8\times 10^{-18}$~W/$\sqrt{\rm Hz}$  &  Assuming AC modulation to cancel $e_n$\\
DC power sensitivity (5-$\sigma$) &  $3.08\times 10^{-20}$~W/(40 g) &  After 2 weeks of DM search (50\% livetime) \\
Energy resolution & $\sigma_E = 12.5$~eV  &  Energy deposition in target absorber\\
Timing resolution & $\sigma_t = 0.68$~s.eV  &  $\sigma_t \approx 6.8\times 10^{-4}$~s at 1~keV\\
\hline
\end{tabular}
\caption{Characteristics of the thermal model simulation shown in Fig.~\ref{fig:NEP_RealDetectors} (right panel).}
\label{tab:parameter_40gDetector}
\end{center}
\end{table}

\end{document}